\documentclass[nohyperref]{article}

\usepackage{microtype}
\usepackage{graphicx}
\usepackage{subfigure}
\usepackage{booktabs} 

\usepackage{authblk}

\usepackage{hyperref}

 \usepackage[accepted]{icml2023}

\usepackage{amsmath}
\usepackage{amssymb}
\usepackage{mathtools}
\usepackage{amsthm}
\usepackage{commath}

\usepackage[capitalize,noabbrev]{cleveref}

\theoremstyle{plain}

\theoremstyle{definition}

\theoremstyle{remark}

\usepackage[textsize=tiny]{todonotes}

\icmltitlerunning{Submission and Formatting Instructions for ICML 2023}

\newcommand{\etal}{\textit{et al}. }

\newcommand\given[1][]{\:#1\vert\:}

\begin{document}
\twocolumn[
\icmltitle{Combination of Multi-Fidelity Data Sources For Uncertainty Quantification: A Lightweight CNN Approach}



\icmlsetsymbol{equal}{*}

\begin{icmlauthorlist}
\icmlauthor{Minghan Chu}{equal,yyy}
\icmlauthor{Weicheng Qian}{equal,comp}

\icmlauthor{Department of Mechanical and Materials Engineering}{yyy}
\icmlauthor{Department of Computer Science}{comp}

\icmlauthor{Queen's University}{yyy}
\icmlauthor{University of Saskatchewwan}{comp}

\icmlauthor{\texttt{17MC93@queensu.ca}}{yyy}
\icmlauthor{\texttt{weicheng.qian@usask.ca}}{comp}

\end{icmlauthorlist}
%
  



\vskip 0.3in
]




\begin{abstract}
Reynolds Averaged Navier Stokes (RANS) modelling is notorious for introducing the model-form uncertainty due to the Boussinesq turbulent viscosity hypothesis. Recently, the eigenspace perturbation method (EPM) has been developed to estimate the RANS model-form uncertainty. This approach estimates model-form uncertainty through injecting perturbations to the predicted Reynolds stress tensor. However, there is a need for a reliable machine learning method for estimating the perturbed amplitude of the Reynolds stress tensor. Machine learning models are often too complex and data intensive for this application. We propose a lightweight convolutional neural network (CNN) approach to learn a correction function for RANS from paired-samples of RANS and DNS simulation results. The CNN learned RANS correction function successfully facilitates the RANS uncertainty quantification (UQ), and our findings suggest that the lightweight CNN approach is effective in combining RANS and DNS simulation results to enrich the existing perturbation method in estimating the RANS UQ more precisely.
\end{abstract}

\section{Introduction}
\label{submission}

Turbulent flows are phenomena that manifest over a wide range of length and time scales. Such length scales vary from micro-meters to kilo-meters. Furthermore these motions are coupled across length scales and exhibit strong interactions. A similar scenario exists for the range of time scales as well. Any numerical simulation of a turbulent flow needs to either explicitly resolve or model these ranges of motions. There are different approaches to such simulations, varying by the ranges of length scales that they resolve and those that they model. For example direct numerical simulation (DNS) computes all the scales of motion, while large eddy simulation (LES) computes only the largest scales while modelling the smaller length scales of motion in the flow. Both DNS and LES methods can yield high fidelity results, while also incurring a considerable increase in computational cost. Reynolds Averaged Navier Stokes (RANS) based simulations use simplified models for all scales. Due to this, RANS-based simulations are relatively computationally inexpensive and are widely used in industry and academia for analyzing turbulent flows. However, most RANS models are often inaccurate when predicting complex flows with Boussinesq turbulent viscosity hypothesis (TVH) adopted \cite{craft1996development}. Therefore, a compromise is to stick with the low-fidelity RANS simulations at low computational cost with uncertainties properly quantified.   

Uncertainty studies can be roughly classified into two categories: 1) aleatory UQ and 2) epistemic UQ. Aleatory uncertainties are introduced in the imprecision of a system \cite{duraisamy2019turbulence}. Studies that have focused on the aleatory uncertainties replace model parameters with random variables \cite{loeven2008airfoil,ahlfeld2017single}, defined a domain region as a random field \cite{dow2015implications,doostan2016bi}, and quantify the irreducible aleatory uncertainties associated with the boundary conditions \cite{pecnik2011assessment}. Epistemic uncertainties are intrinsic in proposed turbulence models, hence known as model-form uncertainty \cite{duraisamy2019turbulence}. The model-form uncertainties have often been ignored in uncertainty studies, while such uncertainties are at a higher-level uncertainty than the aleatory ones \cite{duraisamy2017status}. Less than a decade ago, Iaccarino \emph{et al.} proposed a physics-based eigenspace perturbation approach \cite{emory2013modeling,iaccarino2017eigenspace} to estimate the model-form uncertainty introduced in RANS-based models via sequential perturbations to the amplitude (turbulent kinetic energy), shape (eigenvalues), and orientation (eigenvectors) of the predicted Reynolds stress tensor. The Eigenspace Perturbation Method has been applied with considerable success for Civil engineering applications \cite{gorle2019epistemic}, Aerospace engineering applications \cite{mishra2019uncertainty, mishra2017rans, mishra2019estimating}, Robust and Reliability Based Designs \cite{cook2019optimization, mishra2020design}, besides others. However a central shortcoming of the Eigenspace Perturbation Method is that it always presupposes the worst case scenario and thus leads to overly conservative uncertainty estimates. A central contribution of this investigation is to ensure that the uncertainty estimates are sharp and well calibrated.

\paragraph{Related Works} 
 Data driven approaches like Machine Learning models have found wide application in physics in general and turbulence modelling specifically\cite{duraisamy2019turbulence, ihme2022combustion, brunton2020machine}. Recent machine learning models that have specifically focused on the estimation of RANS model uncertainty with improved accuracy \cite{xiao2016quantifying,wu2016bayesian,xiao2017random,wang2017physics,wang2017comprehensive,wu2018physics,heyse2021estimating,heyse2021data,zeng2022adaptive} are often complex and demand a large size of training data. Complex machine learning models not only require additional computational resources in training but also become less comprehensive to researchers. This hinders the understanding and shrinks the room for improvement in the existing theories. In general, machine learning models in RANS UQ can be classified into two categories: 1) physics-based Bayesian inference and 2) data-driven random forest. There are relatively few studies for reducing the model-form uncertainty using the convolutional neural network (CNN) approach. Besides, most studies have focused on reducing the model-form uncertainty intrinsic to eigenvalues and eigenvectors, except very recently, Chu \textit{et al.} \cite{chu2022model} used polynomial regressions to quantify the model-form uncertainty in turbulence kinetic energy. 
 
 Therefore, the purpose of this paper is to advance the understanding of the performance of CNN approach for reducing the model-form uncertainty introduced in turbulence kinetic energy in RANS simulations with complex flow features such as separation and reattachment. The novelty of this work is the presentation of a lightweight CNN to learn correction function for RANS simulation from both RANS and DNS results. This CNN-based correction function is helpful to the physics-based eigenspace perturbation framework \cite{emory2013modeling,iaccarino2017eigenspace}. Our CNN-based correction function method is non-intrusive, meaning that no modifications need to be made to the RANS-based turbulence models.

\section{Method}

\subsection{Eigenspace Perturbation}

Most RANS models have adopted the Boussinesq turbulent viscosity hypothesis \cite{pope2001turbulent} that assumes Reynolds stresses are proportional to the rate of mean strain:

\begin{equation}
    \left\langle u_i u_j\right\rangle=\frac{2}{3} k \delta_{i j}-2 v_{\mathrm{t}}\left\langle S_{i j}\right\rangle,
\end{equation}

where $k$ is the turbulence kinetic energy, $\delta_{i j}$ is the Kronecker delta, $\nu_{t}$ is the turbulent viscosity, and $\left\langle S_{i j}\right\rangle$ is the rate of mean strain tensor.


The model-form uncertainty introduced in RANS modelling can be quantified via the eigenspace perturbation approach \cite{emory2013modeling,iaccarino2017eigenspace}, the perturbed Reynolds stresses are defined as

\begin{equation}\label{Eq:Rij_perturb}
        \left\langle u_{i} u_{j}\right\rangle^{*}=2 k^{*}\left(\frac{1}{3} \delta_{i j}+v_{i n}^{*} \hat{b}_{n l}^{*} v_{j l}^{*}\right),
\end{equation}

where $k^{*}$ is the perturbed turbulence kinetic energy, $\hat{b}_{k l}^{*}$ is the perturbed eigenvalue matrix, $v_{i j}^{*}$ is perturbed eigenvector matrix.

\begin{equation}\label{Eq:Marker_Mk_Method}
    k^{*} = f^{*}(x,y)
\end{equation}

A better estimated $k^{*}$ can improve UQ. Correction function of RANS is often used to correct RANS towards DNS, resulting a better estimated $k^{*}$.


\subsection{Correction Function for RANS}
Low-fidelity RANS simulations only focus on the mean flow quantities, and hence, are computationally inexpensive. In general RANS simulations are capable of generating satisfactory results for simple shear flows \cite{pope2001turbulent}; however, RANS simulations are inaccurate in predicting complex flow features such as the separation bubble. On the other hand, DNS simulations give high-fidelity results by resolving all scales of fluid motion, while DNS requires tremendous amount of computational resources. Therefore, it is beneficial to correct low-fidelity RANS simulations with the accuracy of high-fidelity DNS. In the present study, we adopt a correction function to correct RANS results towards DNS, namely correction function.

Owing to the importance of RANS model accuracy for engineering design, reliability, and safety, correction function for RANS models has seen a rapid growth of interest in the last few years, such as a linear form \cite{ahlfeld2016multi} and additive and multiplicative form \cite{voet2021hybrid}. 

 For both the RANS and DNS simulation, we can summarize their results as the function of the perturbed turbulence kinetic energy $k$: 
 
\begin{equation}\label{Eq:Marker_Mk_Method}
    k = f(x,y)
\end{equation}

where $x$ and $y$ are coordinates in a two-dimensional computational domain, and $f$ is the mapping from every coordinate $(x, y)$ to $k$, embedded in triples $(x, y, k)$ from simulation results.

Without assuming a specific form, the correction function for RANS is a mapping between two functions:

\begin{equation*}
Z: f^{\mathrm{RANS}}(x,y) \rightarrow f^{\mathrm{DNS}}(x,y)
\end{equation*}

with $k^{\mathrm{DNS}} = f^{\mathrm{DNS}}(x, y)$ and $k^{\mathrm{RANS}} = f^{\mathrm{RANS}}(x, y)$, we can rewrite $Z$ as a mapping $\zeta$ between points that comprises $f^{\mathrm{RANS}}$ and $f^{\mathrm{DNS}}$

\begin{equation*}
\zeta: (x, y, k^{\mathrm{RANS}}) \rightarrow (x, y, k^{\mathrm{DNS}})
\end{equation*}

Consider the model error for RANS and DNS in terms of kinetic energy, we have

\begin{align*}
p^{\mathrm{RANS}}(K_g \given x, y) &= p(k_g = k^{\mathrm{RANS}} \given x, y) \\
p^{\mathrm{DNS}}(K_g \given x, y) &= p(k_g = k^{\mathrm{DNS}} \given x, y)
\end{align*}

where $K_g$ is the unknown ground truth of kinetic energy at $(x, y)$.

Kinetic energy resulted from DNS simulation results $p^{\mathrm{RANS}}$ can be estimated with kinetic energy from RANS simulation $p^{\mathrm{DNS}}$ and its correction function $g$ as

\begin{equation*}
p^{\mathrm{DNS}}(K_g \given x, y) = g(k^{\mathrm{RANS}}, x, y) p(k^{\mathrm{RANS}} \given x, y)
\end{equation*}

Because $k^{\mathrm{DNS}} = f^{\mathrm{DNS}}(x, y)$ and $k^{\mathrm{RANS}} = f^{\mathrm{RANS}}(x, y)$,
at each $x$, we have that $k_x^{\mathrm{DNS}} = f_x^{\mathrm{DNS}}(y)$ and $k_x^{\mathrm{RANS}} = f_x^{\mathrm{RANS}}(y)$, assuming both $f_x^{\mathrm{RANS}}$ and $f_x^{\mathrm{DNS}}$ are continuous, that is, $\forall \epsilon > 0,\allowbreak\, \exists \delta > 0,\allowbreak\, s.t.\, \forall \abs{d} < \delta, \allowbreak \, \abs{f_x(y + d) - f_x(y)} < \epsilon$. We can approximate $g(k^{\mathrm{RANS}}, x, y)$ with $\hat{g}(\mathbf{k}_{x,y,\delta}^{\mathrm{RANS}})$, where $\mathbf{k}_{x,y,\delta}^{\mathrm{RANS}} = [k_{x, y_0}^{\mathrm{RANS}}, k_{x, y_1}^{\mathrm{RANS}}, \cdots]^\top$ and $y_0, y_1, \dots \in [y - \delta, y + \delta]$. In another word, we can learn $\hat{g}$ with paired $(\mathbf{k}_{x,y,\delta}^{\mathrm{RANS}}, \mathbf{k}_{x,y,\delta}^{\mathrm{DNS}})$.

\subsection{CNN-based Correction Function}
\label{sec:methodology-cnn}
We employed a one-dimensional convolutional neural network (1D-CNN) to learn the correction function $\hat{g}$ from paired RANS and DNS simulation estimated kinetic energy $(\mathbf{k}_{x,y,\delta}^{\mathrm{RANS}}, \mathbf{k}_{x,y,\delta}^{\mathrm{DNS}})$. Because our approximated correction function $\hat{g}$ only depends on the neighbor of $k^{\mathrm{RANS}}$ and coordinates $(x, y)$ are only used to group neighbors of $k^{\mathrm{RANS}}$, we grouped simulation data by $x$ and transformed $(y, k)$ at $x$ into $\mathbf{k}_{x,y,\delta}^{\mathrm{RANS}}$ via a rolling window parameterized by window size. Our 1D-CNN has four-layers and in total 86 parameters: a single model for all zones at any $x$ to correct RANS towards DNS.

\section{Experiments Setup and Data Sources}
We experimented our lightweight CNN-based approach to approximate the correction function for RANS on two datasets: the in-house RANS/DNS \cite{zhang2021turbulent,chu2022model} dataset and the public RANS/DNS dataset \cite{voet2021hybrid}. The in-house RANS/DNS dataset \cite{zhang2021turbulent} was obtained by considering the flow around an SD7003 airfoil, with a separation bubble formed on the suction side of the airfoil due to the adverse pressure gradient. The public RANS/DNS dataset \cite{voet2021hybrid} was generated from the two-dimensional channel flow over periodically arranged hills. Similar to the flow over the airfoil, the flow experiences adverse pressure gradient when encountering the curved surface of the hill, which causes the formation of a separation bubble behind the hill. The in-house DNS dataset \cite{zhang2021turbulent} was obtained by considering the flow around an SD7003 airfoil, with a separation bubble formed on the suction side of the airfoil due to the adverse pressure gradient. The public dataset \cite{voet2021hybrid} was generated from DNS flows over periodically arranged hills. Similar to the flow over the airfoil, the flow experiences adverse pressure gradient when encountering the curved surface of the hill, which results in the formation of a separation bubble behind the hill.  

We split $x$-coordinate grouped pairs of $(\mathbf{k}_{x,y,\delta}^{\mathrm{RANS}}, \mathbf{k}_{x,y,\delta}^{\mathrm{DNS}})$ into train set and validation set by their group key $x$. For both the in-house DNS and the public dataset, we choose $x$ at only three positions from the beginning, the middle and the end of all paired $x$ values. For the in-house dataset, $x = 0.4, 0.56, 0.58$; for the public dataset, the $x = 0, 0.046, 0.116, 0.128$. For each dataset, we have a 80\%--20\% split as training--testing dataset. 

\begin{figure*}[h!]
    \centering
    \includegraphics[width=\linewidth]{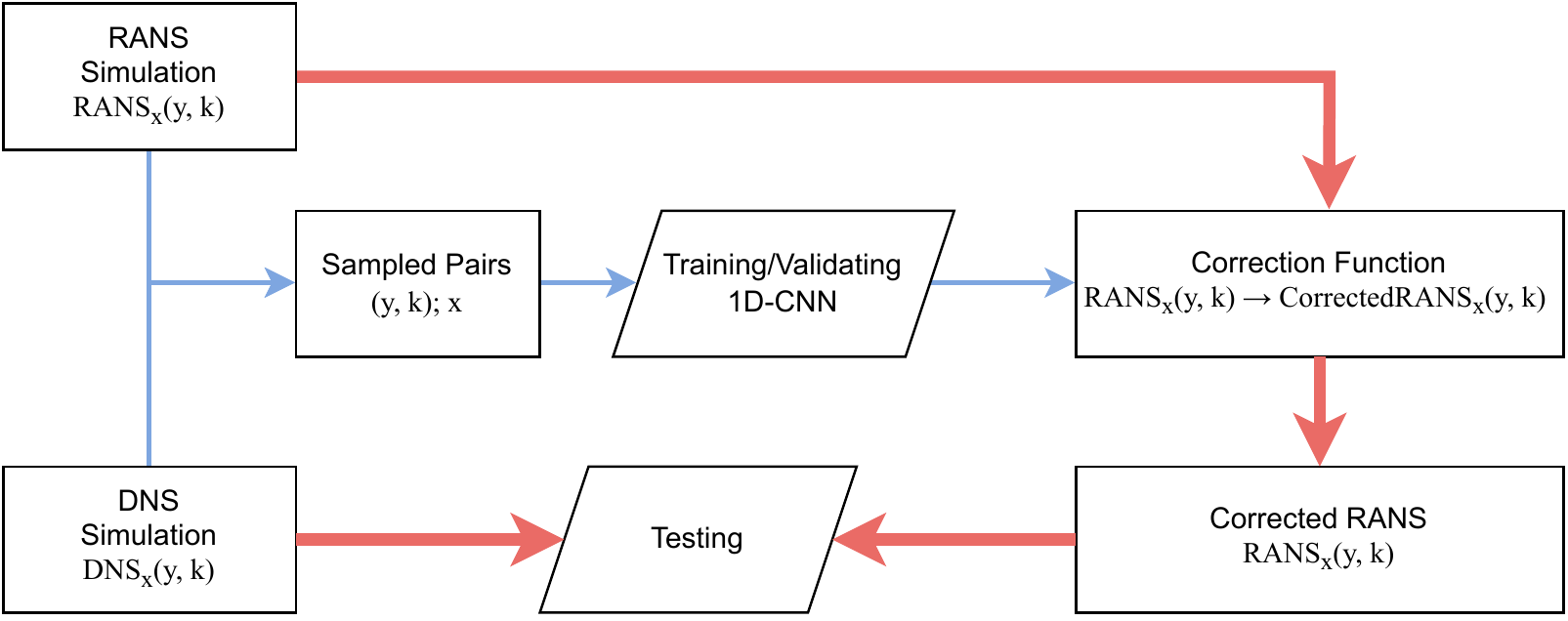}
    \caption{Data-flow Diagram for Experiments. Blue path is the training/validating path, and the red path is the validation path. Thickness of the path indicates the amount of data through the flow.}
    \label{fig:data-flow.pdf}
\end{figure*}

For both datasets, we validated our trained 1D-CNN by comparing the L1 loss of RANS, denoted as $L^1_c(\texttt{rans}) = \abs{ CF^{\mathrm{RANS}}_{k} - CF^{\mathrm{DNS}}_{k} }$ with the L1 loss of 1D-CNN corrected RANS, denoted as $L^1_c(\texttt{pred}) = \abs{ CF^{\mathrm{CNN}}_{k} - CF^{\mathrm{DNS}}_{k} }$.

\section{Results}

\begin{figure*}[!htb]
    \centering
    \includegraphics[width=\linewidth]{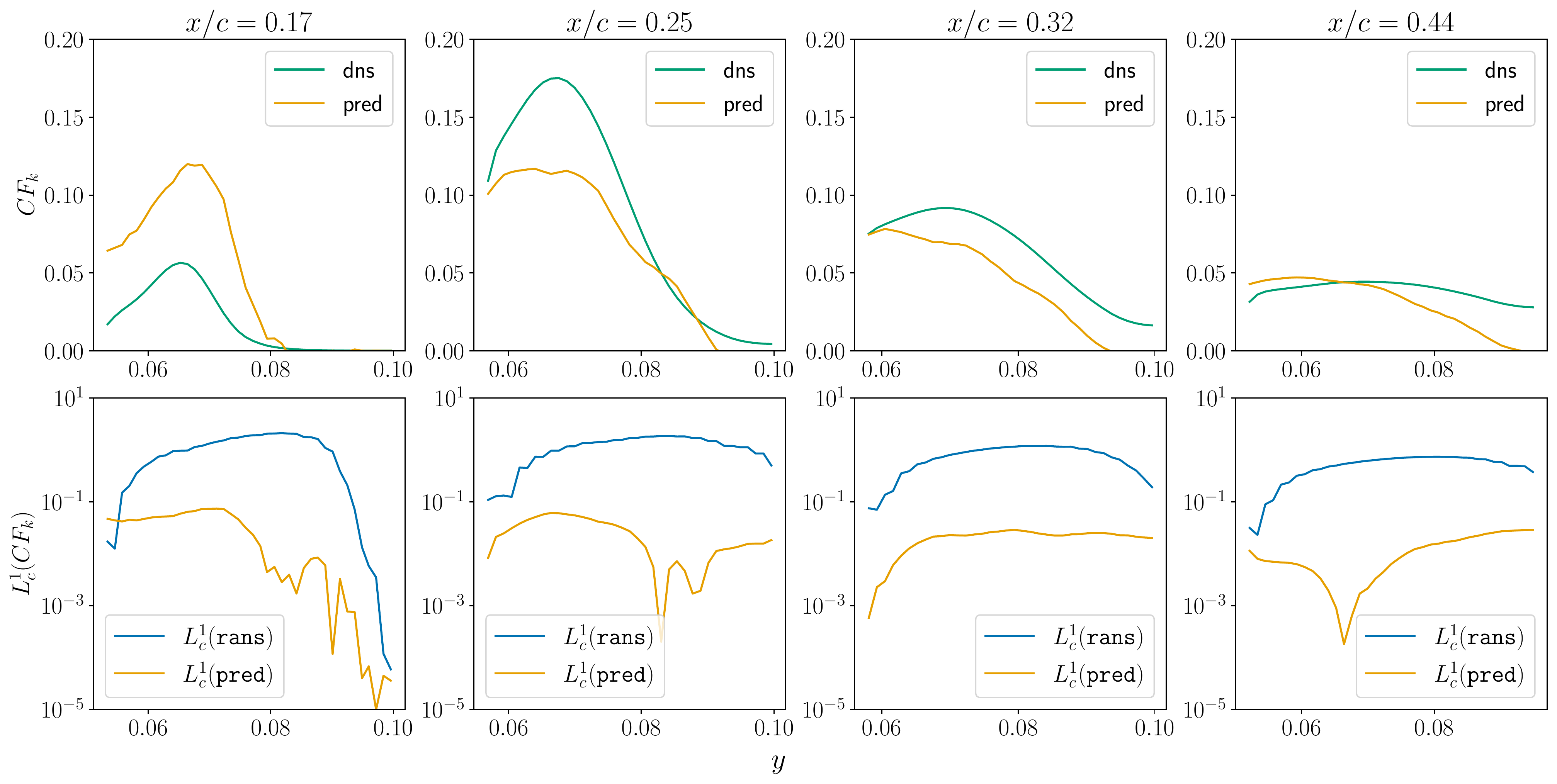}
    \caption{Results for Selig-Donovan 7003 airfoil. First row: CNN corrected DNS (\texttt{pred}) compared with ground truth (\texttt{dns}). Second row: Validation of 1D-CNN by comparing L1 loss between $L^1_c(\texttt{rans})$ and $L^1_c(\texttt{pred})$.}
    \label{fig:cnn-corrected-rans-zhang.pdf}
\end{figure*}

\begin{figure*}[!htb]
    \centering
    \includegraphics[width=\linewidth]{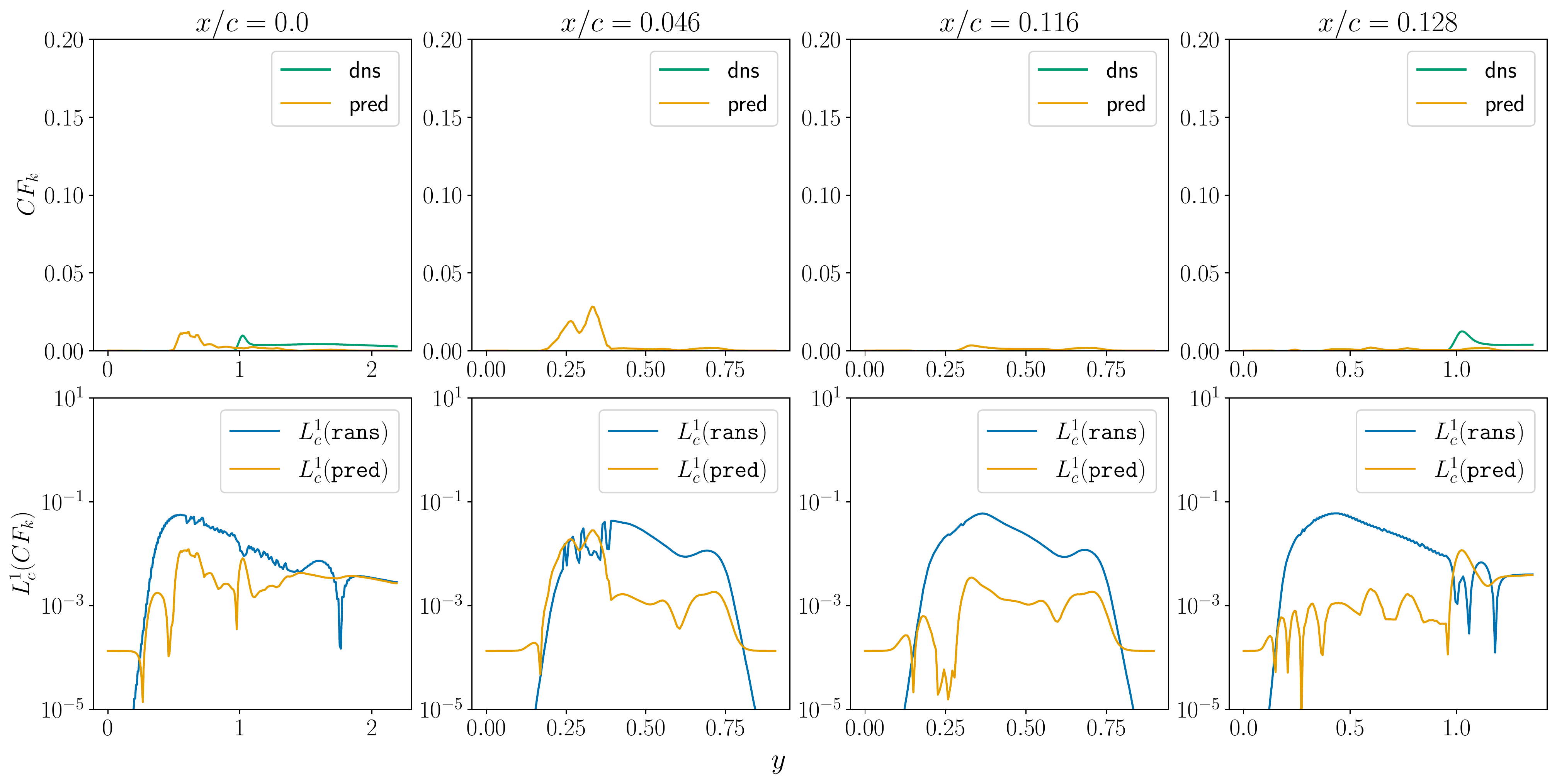}
    \caption{Results for data set from Voet \etal. First row: CNN corrected DNS (\texttt{pred}) compared with ground truth (\texttt{dns}). Second row: Validation of 1D-CNN by comparing L1 loss between $L^1_c(\texttt{rans})$ and $L^1_c(\texttt{pred})$.}
    \label{fig:cnn-corrected-rans-voet.pdf}
\end{figure*}

Our CNN-based correction function is validated at all paired $x$ locations, \Cref{fig:cnn-corrected-rans-zhang.pdf,fig:cnn-corrected-rans-voet.pdf} are four typical $x$ locations for each dataset. For both datasets, our CNN-based correction function results in RANS results located closer to the DNS results at any zone ($x$). The CNN predicted DNS profiles for $k$ shows overall good resemblance to the DNS dataset, although an over-prediction exists at the beginning of the $ab$ zone. 
From the \Cref{fig:cnn-corrected-rans-zhang.pdf}, the series of CNN predicted DNS profiles in the first row are then smoothed with the moving average with a window size of six consecutive estimations. Our CNN predicted DNS profiles resemble the ground truth DNS despite being trained with only a few pairs of RANS and DNS results. From \Cref{fig:cnn-corrected-rans-zhang.pdf}, the discrepancy in general reduces as the flow proceeds further downstream. From the \Cref{fig:cnn-corrected-rans-zhang.pdf}, the second row shows the computed L1 error of the baseline solution and the CNN predicted DNS. It is clear that the L1 error for CNN-based correction function can significantly reduce the L1 error in magnitude compared to that for the original RANS. 


\section{Discussion}
We proposed a CNN approach to approximate the correction function that corrects RANS simulation towards DNS simulation. We further examined our method on two datasets: 1) one flow being considered is over a SD7003 airfoil at $8^\circ$ angle of attack and the Reynolds number based on the cord length of $Re_{c} = 60000$ \cite{zhang2021turbulent}. A laminar separation bubble evolves on the suction side of the airfoil whereby the flow undergoes transition to turbulence, 2) another is generated from the DNS two-dimensional channel flow over periodically arranged hills \cite{voet2021hybrid}. The flow experiences adverse pressure gradient when encountering the curved surface of the hill. It should be noted that a separation bubble occurs for both datasets. The RANS results deviate from the DNS data in both flow scenarios and our CNN-based correction function can significantly reduce the L1 error of RANS- from DNS-simulations.


For both datasets, the CNN-based correction function is trained with paired RANS-DNS simulated turbulence kinetic energy using less than 20\% positions on the $x$-axis, but the trained correction function is effective for the remaining 80\% positions on the $x$-axis. Furthermore, our lightweight CNN model uses the $y$-axis only for grouping RANS-simulated turbulence kinetic energy within a neighbor. The results of our CNN-based correction function suggests that RANS results might be improved by leveraging information embedded in the positions within a close neighbor, which is independent of the absolute coordinates $(x, y)$.

There are relatively few studies for correcting the perturbed turbulence kinetic energy. Very recently, the study of Chu \textit{et al.} \cite{chu2022model} assessed the effect of polynomial regression on the estimation of the perturbed turbulence kinetic energy. Our CNN-based correction method has readily implications on practical applications, such as, to be coupled to the eigenspace perturbation approach of Emory \textit{et al.} \cite{emory2013modeling}. The eigenspace perturbation approach has been implemented within the OpenFOAM framework to construct a marker function for the perturbed turbulence kinetic energy \cite{chu2022model}. Our CNN-based correction method can be used as a new marker function to predict the perturbed turbulence kinetic energy.

\subsection{Application on UQ for airfoil}
Our CNN-based correction function method can be applied to different flow cases to correct RANS towards DNS. In this section, the CNN-based correction function is applied to the SD7003 airfoil case to predict the perturbed turbulence kinetic energy.

\begin{figure*}[t]
         \centering
         \includegraphics[width=\linewidth]{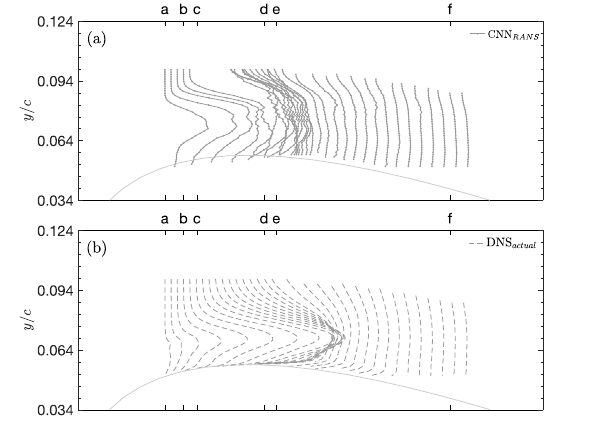}
        \caption{(a) CNN corrected RANS (\texttt{CNN\_{DNS}}) {(solid-dotted lines)} of the normalized perturbed turbulence kinetic energy and (b) ground truth (\texttt{CNN\_{RANS}}) along the suction side of the SD7003 airfoil (geometry depicted by gray line): from left to right are zone $ab$, zone $cd$ and zone $ef$. There are 32 positions on the suction side of the airfoil.}
        \label{fig:CNN_DNS.pdf}
\end{figure*}

The CNN corrected RANS and ground truth profiles for the turbulence kinetic energy normalized with the freestream velocity squared, $k^{*}/U_{\infty}^2$ and $k/U_{\infty}^2$ are shown in Figs. \ref{fig:CNN_DNS.pdf} (a) and (b), respectively. The $k^{*}/U_{\infty}^2$ and $k/U_{\infty}^2$ profiles are equally spaced for the $ab$ and $cd$ zone with $x/c = 0.01$, and a uniform spacing of $x/c = 0.02$ is used for the $ef$ zone. It is clear that the $k^{*}/U_{\infty}^2$ and $k/U_{\infty}^2$ profiles are more densely packed for the $ab$ and $cd$ zone, within which the flow features are complex due to the presence of separation and reattachment. 

From Figs. \ref{fig:CNN_DNS.pdf} (a) and (b), the CNN corrected DNS profiles in general exhibit a similar trend as that for the ground truth dataset, as both profiles show a gradual increase in the $ab$ and $cd$ zone. Then a reduction of the profile is observed further downstream in the $ef$ zone. 

In Fig. \ref{fig:CNN_DNS.pdf} (a), CNN corrected RANS profiles in general increase in magnitude as the flow moves further downstream, which is qualitatively similar to the ground truth profiles. Further, it should be noted that the CNN corrected RANS profiles increase in a somewhat larger magnitude than that for the ground truth in the $ab$ zone. The discrepancy is more than $50\%$ at the beginning of the $ab$ zone and gradually reduces as the flow moves further downstream, which indicates that a better accuracy of our CNN model is yielded further downstream. This behavior becomes more clear for the $cd$ and $ef$ zone. In the region where the end of the $cd$ zone meets the beginning of the $ef$ zone, the ground truth profiles are clustered due to the complex flow feature of the reattachment \cite{chu2022quantification}, as shown in Figs \ref{fig:CNN_DNS.pdf} (b). This clustering behavior is successfully captured by our CNN model, as shown in Fig. \ref{fig:CNN_DNS.pdf} (a). In the $ef$ zone, our CNN model gives overall accurate predictions for the $k^{*}/U_{\infty}^{2}$ profiles, i.e., the CNN corrected RANS profiles and the ground truth profiles are almost identical.

\section{Conclusion}
\label{sec:Conclusion}

To the best of our knowledge, we are among the first to examine the projection from RANS to DNS using the CNN approach. Our experiment results suggest that the CNN-based correction function corrects the RANS predictions for perturbed turbulence kinetic energy towards the in-house DNS data. A projection that can approximate the in-house DNS data reasonably well from RANS might exist independent of $x$. Our methodology can be easily extended to analyze flows over different types of airfoils.

Our findings are subject to following limitations: our CNN model was trained with only two datasets. Future work may include validation with more datasets using different flow cases, e.g., different types of airfoils.  In addition, our CNN-based correction function will be integrated with the eigenspace perturbation framework to result in accurate perturbations and hence improved estimation of RANS UQ.

\nocite{langley00}

\bibliography{rans-dns-cnn-uq}
\bibliographystyle{icml2023}



\end{document}